\documentclass[american,english,aps,manuscript]{revtex4}
\usepackage[T1]{fontenc}
\usepackage[latin9]{inputenc}
\usepackage{babel}
\usepackage{float}
\usepackage{amsmath}
\usepackage{amssymb}
\usepackage{graphicx}
\usepackage{esint}
\usepackage[unicode=true]
 {hyperref}

\makeatletter
\@ifundefined{textcolor}{}
{%
 \definecolor{BLACK}{gray}{0}
 \definecolor{WHITE}{gray}{1}
 \definecolor{RED}{rgb}{1,0,0}
 \definecolor{GREEN}{rgb}{0,1,0}
 \definecolor{BLUE}{rgb}{0,0,1}
 \definecolor{CYAN}{cmyk}{1,0,0,0}
 \definecolor{MAGENTA}{cmyk}{0,1,0,0}
 \definecolor{YELLOW}{cmyk}{0,0,1,0}
}

\makeatother

\begin{document}
\selectlanguage{american}%

\title{A generalized Lieb-Liniger model}

\selectlanguage{english}%

\author{Hagar Veksler and Shmuel Fishman}

\address{Physics Department, Technion- Israel Institute of Technology, Haifa
32000, Israel}
\begin{abstract}
In 1963, Lieb and Liniger solved exactly a one dimensional model of
bosons interacting by a repulsive $\delta$-potential and calculated
the ground state in the thermodynamic limit. In the present work,
we extend this model to a potential of three $\delta$-functions,
one of them is repulsive and the other two are attractive, modeling
some aspects of the interaction between atoms, and present an approximate
solution for a dilute gas. In this limit, for low energy states, the
results are found to be reduced to the ones of an effective Lieb Liniger
model with an effective $\delta$-function of strength $c_{eff}$
and the regime of stability is identified. This may shed light on
some aspects of interacting bosons.
\end{abstract}
\maketitle

\section{introduction}

The physics of Bose gases is a fascinating and complicated field of
research. Since it involves a many-body problem, analytical results
are rare and in some parameter regimes, one can use approximations
to describe experimental systems with very good accuracy. For example,
for a weakly interacting Bose gas, Mean-Field approximation can be
used to reduce the many body Hamiltonian into a one-body non-linear
Schr$\ddot{\text{o}}$dinger Equation, the Gross-Pitaevskii Equation
\cite{Pethick@Smith,Pita_book,PS_review}. In the opposite limit,
a strongly interacting one dimensional Bose gas can be mapped into
a gas of free Fermions (Tonks\textendash{}Girardeau gas, see, for
example, \cite{Girardeau,Olshanii,Bloch_tonks,beta_science}). Exact
solutions in other regimes are highly desired.

Simple models like the Lieb-Liniger (LL) model, that may not have
direct experimental realization, may alert us to unexpected physical
phenomena that are overlooked when ``reasonable approximations''
are made and motivate experiments \cite{beta_prl_exp,Bloch_tonks,beta_science}.
The model introduced in the present work is of this type.

In their seminal work from 1963 \cite{Lieb}, Lieb and Liniger managed
to solve exactly a one dimensional model for interacting bosons. They
considered the Schr$\ddot{\text{o}}$dinger equation for $N$ particles
interacting via a $\delta$-function potential
\begin{equation}
\left[-\frac{\hbar^{2}}{2m}\sum_{j=1}^{N}\frac{\partial^{2}}{\partial x_{j}^{2}}+c\ensuremath{\sum\limits _{\substack{s=1\\
j>s
}
}^{N}}\delta\left(x_{j}-x_{s}\right)\right]\psi=E\psi\label{eq:1D}
\end{equation}
where $x_{j}$ is the coordinate of the $j$-th particle and $c$
is the amplitude of the $\delta$ function. Making a Bethe ansatz
\footnote{For those who are not familiar with Lieb-Liniger solution, we recommend
on lecture notes by Mikhail Zvonarev, \href{http://cmt.harvard.edu/demler/TEACHING/Physics284/LectureZvonarev.pdf}{http://cmt.harvard.edu/demler/TEACHING/Physics284/LectureZvonarev.pdf}
and on the books \cite{bog_book,Tka_book}.%
}
\begin{equation}
\psi\left(x_{1},...,x_{N}\right)\propto\sum_{P}\left(-1\right)^{\left[P\right]}\exp\left\{ i\sum_{n=1}^{N}x_{n}k_{P_{n}}\right\} \prod_{j>s}\left[k_{P_{j}}-k_{P_{s}}-\frac{imc}{\hbar^{2}}\mathrm{sign}\left(x_{j}-x_{s}\right)\right],\label{eq:psi_lieb}
\end{equation}
where $k_{P_{n}}$ are the $k$ vectors obtained by the permutation
$P$ (where $\left[P\right]$ is its parity) of the set $k_{1},\ldots,k_{N}$.
Lieb and Liniger wrote Bethe ansatz equations for the $k$'s by imposing
periodic boundary conditions on a ring of length $L$ \cite{bog_book,Tka_book},
\begin{equation}
\exp\left\{ ik_{j}L\right\} =\prod_{h\neq j}^{N}\frac{\hbar^{2}\left(k_{j}-k_{h}\right)+imc}{\hbar^{2}\left(k_{j}-k_{h}\right)-imc}=\prod_{h\neq j}^{N}\frac{1+\frac{imc}{\hbar^{2}\left(k_{j}-k_{h}\right)}}{1-\frac{imc}{\hbar^{2}\left(k_{j}-k_{h}\right)}}.\label{eq:bet_lieb}
\end{equation}
These $N$ coupled equations are solved numerically and the energy
\begin{equation}
E=\frac{\hbar^{2}}{2m}\sum_{j=1}^{N}k_{j}^{2}
\end{equation}
is calculated for the ground state and the excitations \cite{Lieb,Liev_Spect,zoran}.

In the present work, we study a simple model which takes into account
the range of inter-particle interactions without giving up the mathematical
simplicity. It is a generalization of the LL model \cite{Lieb} where
in addition to the repulsion there is also attraction. It is defined
by the Schr$\ddot{\text{o}}$dinger equation for $N$ interacting
particles of mass $m$,
\begin{equation}
\left[-\frac{\hbar^{2}}{2m}\sum_{j=1}^{N}\frac{\partial^{2}}{\partial x_{j}^{2}}+c_{0}\ensuremath{\sum\limits _{\substack{s=1\\
j>s
}
}^{N}}\delta\left(x_{j}-x_{s}\right)+c_{l}\ensuremath{\sum\limits _{\substack{s=1\\
j>s
}
}^{N}}\delta\left(x_{j}-x_{s}-l\right)+c_{l}\ensuremath{\sum\limits _{\substack{s=1\\
j>s
}
}^{N}}\delta\left(x_{j}-x_{s}+l\right)\right]\psi=E\psi\label{eq:3D_shrod}
\end{equation}
where the inter-particle interaction is modeled as a sum of three
$\delta$-functions: The central one is repulsive $\left(c_{0}>0\right)$
while the peripheral ones are attractive $\left(c_{l}<0\right)$.
This model is inspired by the Van-der-Waals potential which has repulsive
and attractive regimes. By adjusting the parameters $c_{0},c_{l},l$
of (\ref{eq:3D_shrod}), one can model scattering from many inter-particle
potentials \cite{3_Del_long,3_del_PRL,Ketterle&us}.

In section II, we present Bethe ansatz equations for two bosons interacting
via three $\delta$-functions interaction potential and in section
III an approximation is introduced, that allows to extend the LL Bethe
ansatz equations to an arbitrary number of particles. The ground state
solution for the approximate equations is found in section IV. Section
V specifies the parameters of the regime where the gas is stable.
The results and their experimental relevance are discussed in section
VI.

\section{Bethe ansatz equations for two bosons interacting via three $\delta$-functions
interaction potential}

We start by writing Bethe ansatz equations for a simple case where
there are only two bosons. In this case, the equations are intuitive. 

Consider two bosons of mass $m$ trapped on a ring of length $L$
and interact according to (\ref{eq:3D_shrod}). It is convenient to
write the wave function $\psi$ in terms of center of mass coordinate,
$r_{1}=\left(x_{1}+x_{2}\right)/2$ and relative motion coordinate,
$r_{2}=\left(x_{1}-x_{2}\right)/2$,
\begin{equation}
\psi\left(r_{1},r_{2}\right)=\frac{1}{\sqrt{L}}e^{i\tilde{k}_{1}r_{1}}\phi\left(r_{2}\right)\label{eq:psi6}
\end{equation}
where $\tilde{k_{1}}=2\pi n/L$ and $n$ is an integer so that periodic
boundary conditions are satisfied. At the center of mass frame of
reference, $\tilde{k}_{1}=0$ and the wavefunction of the relative
motion, $\phi\left(r_{2}\right)$, satisfies the Schr$\ddot{\text{o}}$dinger
equation
\begin{equation}
\left[-\frac{\hbar^{2}}{4m}\frac{\partial^{2}}{\partial r_{2}^{2}}+\frac{1}{2}c_{0}\delta\left(r_{2}\right)+\frac{1}{2}c_{l}\delta\left(r_{2}-l/2\right)+\frac{1}{2}c_{l}\delta\left(r_{2}+l/2\right)\right]\phi\left(r_{2}\right)=E\phi\left(r_{2}\right)\label{eq:3D_shrod-1}
\end{equation}
which can be written also as
\begin{equation}
\left[-\frac{\hbar^{2}}{2m}\frac{\partial^{2}}{\partial r_{2}^{2}}+c_{0}\delta\left(r_{2}\right)+c_{l}\delta\left(r_{2}-l/2\right)+c_{l}\delta\left(r_{2}+l/2\right)\right]\phi\left(r_{2}\right)=2E\phi\left(r_{2}\right).\label{eq:3D_shrod-1-1}
\end{equation}
As usual in such cases, the wave function takes a different functional
form in each of the four intervals $\left[-\frac{L}{4},-\frac{l}{2}\right]$,
$\left[-\frac{l}{2},0\right]$, $\left[0,\frac{l}{2}\right]$ and
$\left[\frac{l}{2},\frac{L}{4}\right]$. The result for $\phi\left(r_{2}\right)$
is
\begin{eqnarray}
\phi\left(r_{2}\right) & = & C\cos\left(\tilde{k}_{2}r_{2}\right)+iC\left\{ \mathrm{sign}\left(r_{2}\right)Q_{0}\sin\left(\tilde{k}_{2}r_{2}\right)+\right.\label{eq:phi2}\\
 &  & \left.Q_{l}\mathrm{sign}\left(r_{2}-l/2\right)\sin\left[\tilde{k}_{2}\left(r_{2}-l/2\right)\right]+Q_{l}\mathrm{sign}\left(r_{2}+l/2\right)\sin\left[\tilde{k}_{2}\left(r_{2}+l/2\right)\right]\right\} \nonumber 
\end{eqnarray}
where $C$ is a normalization constant while $Q_{0}$ and $Q_{l}$
should be determined. They are easily determined for the $\delta$-function
interaction since the jump of the derivative at the locations of the
$\delta$-function satisfies
\begin{equation}
\Delta\phi^{\prime}\left(r_{2}^{*}\right)\equiv\left.\frac{d\phi\left(r_{2}\right)}{dr_{2}}\right|_{r_{2}^{*}+0^{+}}-\left.\frac{d\phi\left(r_{2}\right)}{dr_{2}}\right|_{r_{2}^{*}+0^{-}}=\frac{2mc^{*}}{\hbar^{2}}\phi\left(r_{2}^{*}\right)
\end{equation}
where $r_{2}^{*}=0,\, c^{*}=c_{0}$ or $r_{2}^{*}=l/2,\, c^{*}=c_{l}$.
This results in two equations for $Q_{0}$ and $Q_{l}$
\begin{equation}
i\tilde{k}_{2}Q_{0}\left(\tilde{k}_{2}\right)=\frac{mc_{0}}{\hbar^{2}}\left\{ 1+2iQ_{l}\sin\left(\tilde{k}_{2}l/2\right)\right\} \label{eq:Q0_dev}
\end{equation}
and
\begin{equation}
i\tilde{k}_{2}Q_{l}=\frac{mc_{l}}{\hbar^{2}}\left\{ \cos\left(\tilde{k}_{2}l/2\right)+i\left[Q_{0}\sin\left(\tilde{k}_{2}l/2\right)+Q_{l}\sin\left(\tilde{k}_{2}l\right)\right]\right\} ,\label{eq:Ql_dev}
\end{equation}
leading to
\begin{equation}
Q_{l}\left(\tilde{k}_{2}\right)=-\frac{i\tilde{k}_{2}\hbar^{2}mc_{l}\left[\cos\left(\tilde{k}_{2}l/2\right)+\frac{mc_{0}}{\tilde{k}_{2}\hbar^{2}}\sin\left(\tilde{k}_{2}l/2\right)\right]}{\tilde{k}_{2}^{2}\hbar^{4}-2m^{2}c_{0}c_{l}\sin^{2}\left(\tilde{k}_{2}l/2\right)-\tilde{k}_{2}\hbar^{2}mc_{l}\sin\left(\tilde{k}_{2}l\right)}.\label{eq:Q_l}
\end{equation}
$\tilde{k}_{2}$ should be determined to ensure periodic boundary
conditions $\phi^{\prime}\left(r_{2}\right)=\phi^{\prime}\left(r_{2}+L/2\right)$.
In addition, $\phi\left(r_{2}\right)=\phi\left(-r_{2}\right)$ so
that $\phi^{\prime}\left(r_{2}\right)=-\phi^{\prime}\left(-r_{2}\right)$.
Therefore, in particular, $\left.\phi^{\prime}\left(r_{2}\right)\right|_{r_{2}=L/4}$
must vanish, leading to 
\begin{equation}
e^{i\tilde{k}_{2}L/2}=\frac{1-Q_{0}\left(\tilde{k}_{2}\right)-2Q_{l}\left(\tilde{k}_{2}\right)\cos\left(\tilde{k}_{2}l/2\right)}{1+Q_{0}\left(\tilde{k}_{2}\right)+2Q_{l}\left(\tilde{k}_{2}\right)\cos\left(\tilde{k}_{2}l/2\right)}.\label{eq:bethe_tilde_k}
\end{equation}

Now, we return to coordinates $x_{1},x_{2}$. For this purpose, we
use the relations: $k_{1}=\left(\tilde{k}_{1}+\tilde{k}_{2}\right)/2$
, $k_{2}=\left(\tilde{k}_{1}-\tilde{k}_{2}\right)/2$ , $r_{1}=\left(x_{1}+x_{2}\right)/2$
and $r_{2}=\left(x_{1}-x_{2}\right)/2$ resulting in
\begin{eqnarray}
\tilde{k}_{1}r_{1}+\tilde{k}_{2}r_{2} & = & k_{1}x_{1}+k_{2}x_{2}
\end{eqnarray}
and
\begin{eqnarray}
\tilde{k}_{1}r_{1}-\tilde{k}_{2}r_{2} & = & k_{2}x_{1}+k_{1}x_{2}.
\end{eqnarray}
The function $\psi$ of (\ref{eq:psi6}) takes the form 
\begin{eqnarray}
\psi\left(x_{1},x_{2}\right) & = & C\left[e^{i\left(k_{1}x_{1}+k_{2}x_{2}\right)}+e^{i\left(k_{2}x_{1}+k_{1}x_{2}\right)}\right]\label{eq:2body_psi}\\
 &  & +C\mathrm{sign}\left(x_{1}-x_{2}\right)Q_{0}\left(k_{1}-k_{2}\right)\left(e^{i\left(k_{1}x_{1}+k_{2}x_{2}\right)}-e^{i\left(k_{2}x_{1}+k_{1}x_{2}\right)}\right)\nonumber \\
 &  & +CQ_{l}\left(k_{1}-k_{2}\right)\mathrm{sign}\left(x_{1}-x_{2}-l\right)\left[e^{i\left(k_{1}x_{1}+k_{2}x_{2}\right)}e^{-i\left(k_{1}-k_{2}\right)l/2}-e^{i\left(k_{2}x_{1}+k_{1}x_{2}\right)}e^{i\left(k_{1}-k_{2}\right)l/2}\right]\nonumber \\
 &  & +CQ_{l}\left(k_{1}-k_{2}\right)\mathrm{sign}\left(x_{1}-x_{2}+l\right)\left[e^{i\left(k_{1}x_{1}+k_{2}x_{2}\right)}e^{i\left(k_{1}-k_{2}\right)l/2}-e^{i\left(k_{2}x_{1}+k_{1}x_{2}\right)}e^{-i\left(k_{1}-k_{2}\right)l/2}\right].\nonumber 
\end{eqnarray}
 The periodic boundary condition $\psi\left(x_{2}+\frac{L}{2},x_{2}\right)=\psi\left(x_{2}-\frac{L}{2},x_{2}\right)$
results in 
\begin{equation}
e^{ik_{1}L}=\frac{1-Q_{0}\left(k_{1}-k_{2}\right)-2Q_{l}\left(k_{1}-k_{2}\right)\cos\left(\left(k_{1}-k_{2}\right)l/2\right)}{1+Q_{0}\left(k_{1}-k_{2}\right)+2Q_{l}\left(k_{1}-k_{2}\right)\cos\left(\left(k_{1}-k_{2}\right)l/2\right)}\label{eq:bet12}
\end{equation}
and
\begin{equation}
e^{ik_{2}L}=\frac{1+Q_{0}\left(k_{1}-k_{2}\right)+2Q_{l}\left(k_{1}-k_{2}\right)\cos\left(\left(k_{1}-k_{2}\right)l/2\right)}{1-Q_{0}\left(k_{1}-k_{2}\right)-2Q_{l}\left(k_{1}-k_{2}\right)\cos\left(\left(k_{1}-k_{2}\right)l/2\right)}\label{eq:bet21}
\end{equation}
 which are identical to (\ref{eq:bethe_tilde_k}) (under the assumption
$\tilde{k}_{1}=0$, namely, in the center of mass frame of reference).
In the derivation we used the fact that $x_{1}-x_{2}\rightarrow x_{1}-x_{2}+L$
involves rotation around the circle and consequently all the signs
are changed.

\section{Approximate Bethe ansatz equations for an arbitrary number of bosons}

The two particle solution cannot be simply generalized to an arbitrary
number of particles since for small interparticle distances,
\begin{equation}
\left|x_{j}-x_{s}\right|<l,\label{eq:xl}
\end{equation}
the sign function in equation corresponding to (\ref{eq:2body_psi})
varies substantially. For small $l$, the effect of the regime (\ref{eq:xl})
may be negligible as demonstrated in what follows. This is reasonable
for a dilute gas where $l\ll L/N$. In such a situation, the LL solution
is valid with the replacement $\frac{imc}{\hbar^{2}\left(k_{j}-k_{s}\right)}=Q_{0}+2Q_{l}\cos\left(\left(k_{j}-k_{s}\right)l/2\right)$,
leading to

\begin{eqnarray}
\psi\left(x_{1},...,x_{N}\right) & =C & \sum_{P}\left[\exp\left\{ i\sum_{n=1}^{N}x_{n}k_{P_{n}}\right\} \right.\label{eq:psi_N}\\
 &  & \prod_{j>s}\left\{ 1+\left(Q_{0}\left(k_{j}-k_{s}\right)+2Q_{l}\left(k_{j}-k_{s}\right)\cos\left(\left(k_{j}-k_{s}\right)l/2\right)\right)\mathrm{sign}\left(x_{j}-x_{s}\right)\right\} \nonumber 
\end{eqnarray}
and
\begin{equation}
e^{ik_{j}L}=\prod_{s\neq j}\frac{1-Q_{0}\left(k_{j}-k_{s}\right)-2Q_{l}\left(k_{j}-k_{s}\right)\cos\left(\left(k_{j}-k_{s}\right)l/2\right)}{1+Q_{0}\left(k_{j}-k_{s}\right)+2Q_{l}\left(k_{j}-k_{s}\right)\cos\left(\left(k_{j}-k_{s}\right)l/2\right)}.\label{eq:betN}
\end{equation}
 The $k_{j}$ are distinct, namely, the wave function vanishes if
$k_{j}=k_{s}$ for $s\neq j$ as was shown in the original work of
LL \cite{Lieb}. 

In the region where inequalities (\ref{eq:xl}) are not satisfied
for any of the particle pairs, the $\mathrm{sign}$ functions in (\ref{eq:phi2})
are all equal. Therefore, in this regime, (\ref{eq:psi_N}) is a solution
with the spectrum (\ref{eq:betN}). There is a Hamiltonian that is
different from the original one, for which (\ref{eq:psi_N}) and (\ref{eq:betN})
are eigenfunctions and eigenvalues even if some of the inequalities
(\ref{eq:xl}) are satisfied. It is just defined by the eigenfunctions
and eigenvalues. For $l=0$, this Hamiltonian and the original one
are identical. If the spectrum and the $L^{2}$-norm of the eigenfunctions
are continuous in $l$, the relative difference in the spectrum and
the wavefunctions (in the $L^{2}$-norm) goes to zero in the limit
$l\rightarrow0$. If they are also differentiable as a function of
$l$, then the relative difference behaves as $Nl/L$. 

We show that for the low energy states, the 3-$\delta$ function system
can be replaced by a system with one $\delta$-function of strength
$c_{eff}$.

We assume 
\begin{equation}
\left(k_{j}-k_{s}\right)l\ll1\label{eq:hilazon}
\end{equation}
 for all wave vectors $k_{j}$. In section V, we show that this limit
is relevant for the ground state and low-lying excitations of a dilute
gas since $k_{max}\leq\mathrm{\left(const\right)}\frac{N}{L}$ is
small. In the leading order in $k_{j}l$,
\begin{eqnarray}
Q_{0}+2Q_{l}\cos\left(\left(k_{j}-k_{h}\right)l/2\right) & \approx & -i\frac{m}{\hbar^{2}\tilde{k}_{2}}\left\{ c_{0}+2c_{l}+\frac{\frac{mc_{l}l}{\hbar^{2}}\left[2c_{0}+2c_{l}+\frac{mc_{0}c_{l}l}{\hbar^{2}}+\frac{mc_{0}^{2}l}{2\hbar^{2}}\right]}{\left[1-\frac{m^{2}c_{0}c_{l}l^{2}}{2\hbar^{4}}-\frac{mc_{l}l}{\hbar^{2}}\right]}\right\} ,\label{eq:Q_eff}
\end{eqnarray}
the error is of the order $Nl/L$. Comparing (\ref{eq:betN}) with
(\ref{eq:bet_lieb}), one finds that for small $k_{j}l$, the behavior
of the present problem is indeed similar to the one found for one
$\delta$-function potential of strength
\begin{equation}
c_{eff}=c_{0}+2c_{l}+\frac{\frac{mc_{l}l}{\hbar^{2}}\left[2c_{0}+2c_{l}+\frac{mc_{0}c_{l}l}{\hbar^{2}}+\frac{mc_{0}^{2}l}{2\hbar^{2}}\right]}{\left[1-\frac{m^{2}c_{0}c_{l}l^{2}}{2\hbar^{4}}-\frac{mc_{l}l}{\hbar^{2}}\right]},\label{eq:c_eff}
\end{equation}
in the leading order in $k_{j}l$. Eq. (\ref{eq:c_eff}) is the main
result of the present work, and it enables one to understand the physics
of the three $\delta$-functions interaction in terms of the one $\delta$-function
interaction. Of particular interest are situations where $c_{eff}$
is very different from $c_{0}+2c_{l}$ (the total strength of interactions).
In order to find such situations, we define the parameters $r=c_{l}/c_{0}$
and $x=mc_{0}l/\hbar^{2}$ and rewrite (\ref{eq:c_eff}) as
\begin{eqnarray}
\frac{c_{eff}}{c_{0}+2c_{l}} & = & 1+\frac{rx\left(2+2r+rx+\frac{1}{2}x\right)}{\left(1+2r\right)\left(1-\frac{1}{2}rx^{2}-rx\right)}\label{eq:c_effg}\\
 & = & \frac{rx+1+2r}{\left(1+2r\right)\left(1-\frac{1}{2}rx^{2}-rx\right)}\nonumber 
\end{eqnarray}
For weak interactions, ($x\ll1$ and $rx\ll1$ ),
\begin{equation}
\frac{c_{eff}}{c_{0}+2c_{l}}\approx1.
\end{equation}
However, for very strong interactions ($x\rightarrow\infty$),
\begin{equation}
\frac{c_{eff}}{c_{0}+2c_{l}}\approx\frac{-2}{\left(1+2r\right)x}\rightarrow0^{-}.\label{eq:c_eff0}
\end{equation}
This is a surprising result. It is instructive to analyze the behavior
of $c_{eff}/\left(c_{0}+2c_{l}\right)$, Eq. (\ref{eq:c_effg}), as
a function of $x$. We are interested in the regime $x>0$ and $-0.5<r<0$.
At $x=0$, the derivative of (\ref{eq:c_effg}) is negative and therefore
the function decreases. At 
\begin{equation}
x_{0}=-\frac{\left(1+2r\right)}{r}\label{eq:x_0}
\end{equation}
it turns out that $c_{eff}=0$ (even though $c_{0}+2c_{l}\neq0$).
Higher values of $x$ result in negative values of $c_{eff}$, namely,
the effective interaction is attractive (even though $c_{0}+2c_{l}>0$).
Schematic description of $c_{eff}/\left(c_{0}+2c_{l}\right)$ is given
in Fig. 1.

\selectlanguage{american}%
\begin{figure}[H]
\selectlanguage{english}%
\includegraphics[scale=0.6]{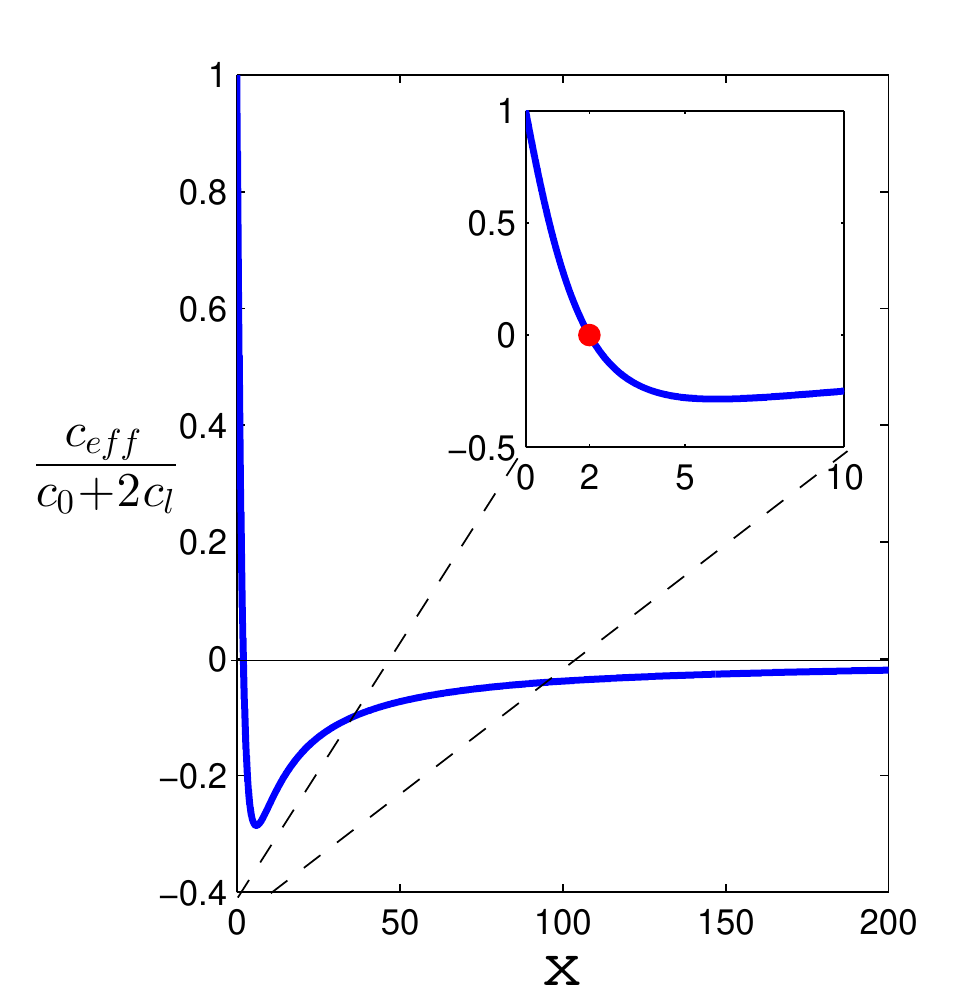}

\selectlanguage{american}%
\caption{\selectlanguage{english}%
(Color online) Schematic description of $c_{eff}/\left(c_{0}+2c_{l}\right)$
as a function of $x$ for $r=c_{l}/c_{0}=-0.25$. The inset expands
the region where $c_{eff}$ changes its sign and the red dot is $\left(x_{0},0\right)$\foreignlanguage{american}{.}\selectlanguage{american}%
}
\end{figure}

\selectlanguage{english}%
The result $c_{eff}=0$ at $x=x_{0}$ is verified numerically (see
Fig. 2) and will be discussed in what follows. In the two particle
case it is exact. For a related result see \cite{3_Del_long,3_del_PRL}.

\section{Ground state energy}

In the previous section, we derived the approximate Bethe ansatz equations
(\ref{eq:betN}) for $N$ bosons interacting by a three $\delta$-function
potential (\ref{eq:3D_shrod}). The solution for these $N$ coupled
equations, $\left(k_{1},k_{2},\ldots,k_{N}\right)$, can be used to
calculate the energy of the gas
\begin{equation}
E=\frac{\hbar^{2}}{2m}\sum_{j=1}^{N}k_{j}^{2}.\label{eq:E}
\end{equation}
In the ground state, $\left|k_{j}\right|$ are minimal (but yet $k_{j}$
are different, as in the original work of LL \cite{Lieb}).

Lieb and Liniger \cite{Lieb} managed to calculate the ground state
energy in the thermodynamic limit ($N\rightarrow\infty$) by solving
only two coupled integral equations (\ref{eq:fermi}) and (\ref{eq:ru})
(instead of $N$ equations of the form (\ref{eq:betN})). Here, we
obtain similar equations by using the logarithmic form of (\ref{eq:betN}),
\begin{equation}
G\left(k_{j}\right)\equiv k_{j}L+\sum_{s\neq j}\theta\left(k_{j}-k_{s}\right)=2\pi\left(n_{j}-\frac{N+1}{2}\right)\label{eq:k_n}
\end{equation}
where
\begin{equation}
\theta\left(k\right)=i\ln\left[\frac{Q_{0}\left(k\right)+2Q_{l}\left(k\right)\cos\left(kl/2\right)-1}{Q_{0}\left(k\right)+2Q_{l}\left(k\right)\cos\left(kl/2\right)+1}\right].\label{eq:theta}
\end{equation}
We see that if $l=0$, the ground state corresponds to the choice
$n_{j}=j$, $\left(j=1,\ldots,N\right)$. This is true also for $l\neq0$,
as long as $\theta$ is a monotonic increasing function of $k$. To
see this, assume $k_{j}>k_{m}$, then, by monotonicity of $\theta$,
$\theta\left(k_{j}-k_{s}\right)>\theta\left(k_{m}-k_{s}\right)$ for
all $s$, therefore $G\left(k_{j}\right)>G\left(k_{m}\right)$ and
$G\left(k_{j}\right)$ is monotonic. Since $\theta$ is an odd function,
$G\left(k_{j}=0\right)=0$. The $k_{j}$ for the ground state are
the smallest possible in absolute value, hence, we choose $n_{j}=j$
for the ground state. Therefore, in the monotonic regime,
\begin{equation}
L\left(k_{j+1}-k_{j}\right)+\left(k_{j+1}-k_{j}\right)\sum_{s\neq j}\theta^{\prime}\left(k_{j}-k_{s}\right)=2\pi\label{eq:29}
\end{equation}
where $\theta^{\prime}\left(k\right)\equiv\partial\theta\left(k\right)/\partial k$
and $k_{j}$ and $k_{j+1}$ are adjacent wave numbers. Typically,
$\theta$ is monotonic and (\ref{eq:29}) is justified at the regime
where (\ref{eq:hilazon}) holds (see Sec. V for more details). The
density of states per unit length in $k$ space, is defined as
\begin{equation}
\rho\left(k_{j}\right)=\frac{1}{L\left(k_{j+1}-k_{j}\right)}\label{eq:ru-1}
\end{equation}
and satisfies
\begin{equation}
\int_{-\Lambda}^{\Lambda}dk\rho\left(k\right)=\frac{N}{L}.\label{eq:fermi}
\end{equation}
It is used to write (\ref{eq:29}) in the form
\begin{equation}
\rho\left(k\right)-\frac{1}{2\pi}\int_{-\Lambda}^{\Lambda}dq\rho\left(q\right)\theta^{\prime}\left(k-q\right)=\frac{1}{2\pi}.\label{eq:ru}
\end{equation}
Here, $\Lambda$ is the Fermi momentum (this should not be confused
with fermionic systems!). The ground state energy (\ref{eq:E}) is
\begin{equation}
E_{0}=\frac{\hbar^{2}L}{2m}\int_{-\Lambda}^{\Lambda}dk\rho\left(k\right)k^{2}.\label{eq:E_k}
\end{equation}

In order to solve Eqs.(\ref{eq:fermi}) and (\ref{eq:ru}), we change
into dimensionless variables:
\begin{equation}
z=\frac{k}{\Lambda},\quad\alpha_{0}=\frac{c_{0}m}{\Lambda\hbar^{2}},\quad\alpha_{l}=\frac{c_{l}m}{\Lambda\hbar^{2}},\quad\gamma_{0}=\frac{c_{0}mL}{\hbar^{2}N},\quad\gamma_{l}=\frac{c_{l}mL}{\hbar^{2}N},\quad d=\Lambda l.\label{eq:dim_less}
\end{equation}
In these variables, $\eta(z)=\theta\left(\Lambda z\right)$, the density
of states is $g\left(z\right)=\rho\left(\Lambda z\right)$, and Eqs.
(\ref{eq:fermi}), (\ref{eq:ru}) and (\ref{eq:E_k}) are, respectively
\cite{Lieb},
\begin{equation}
\gamma_{0}\int_{-1}^{1}dzg\left(z\right)=\alpha_{0},\label{eq:Li1}
\end{equation}
\begin{equation}
g\left(z\right)-\frac{1}{2\pi}\int_{-1}^{1}dy\eta^{\prime}\left(y-z\right)g\left(y\right)=\frac{1}{2\pi}\label{eq:Li2}
\end{equation}
and
\begin{equation}
e\equiv\frac{2mE_{0}L^{2}}{\hbar^{2}N{}^{3}}=\frac{\gamma_{0}^{3}}{\alpha_{0}^{3}}\int_{-1}^{1}dyg\left(y\right)y^{2}.\label{eq:dle}
\end{equation}
How does one solve Eqs. (\ref{eq:Li1}) and (\ref{eq:Li2})? First,
it is necessary to choose values for $\alpha_{0},\alpha_{l}$ and
$d$. These values are related to the parameters of the Hamiltonian
via the Fermi momentum $\Lambda$ which is unknown at this stage.
One should only keep in mind that $\alpha_{l}/\alpha_{0}=c_{l}/c_{0}$
and therefore the ratio $\alpha_{l}/\alpha_{0}$ does reflect the
ratio between attraction and repulsion in the Hamiltonian. The integral
equation (\ref{eq:Li2}) (with parameters $\alpha_{0},\alpha_{l}$
and $d$) can be solved numerically. The solution, $g\left(z\right)$,
should be substituted in (\ref{eq:Li1}) in order to find $\gamma_{0}$.
By repeating the above scheme for different parameters, it is possible
to plot the dimensionless energy $e$ as a function of the dimensionless
interaction strengths $\gamma_{0}$ and $\gamma_{l}$ and the dimensionless
length $d$. for small values of $d$, the energy $e$ depends only
on the effective strength of interaction, that is (\ref{eq:c_eff})
in dimensionless units,
\begin{equation}
\gamma_{eff}=\gamma_{0}+2\gamma_{l}+\frac{\alpha_{l}d\left(2\gamma_{0}+2\gamma_{l}+\alpha_{l}d\gamma_{0}+\frac{1}{2}\alpha_{0}d\gamma_{0}\right)}{1-\frac{1}{2}\alpha_{0}\alpha_{l}d^{2}-\alpha_{l}d}.\label{eq:gam_eff}
\end{equation}
The significance of this effective strength of interaction is demonstrated
in figure 2. In this figure, we present the solutions $e\left(\alpha_{0},\alpha_{l},d\right)$
that were calculated by solving (\ref{eq:Li1}), (\ref{eq:Li2}) and
(\ref{eq:dle}). In Fig. 2(a), the energy is plotted as a function
of the total interaction strength $\gamma_{0}+2\gamma_{l}$ and different
choices of $d$ are represented by different colors. It is clear that
the effect of $d$ is not negligible. Even at the regime $d\ll1$,
it is evident that the value of $l$ has a strong effect on the ground
state energy. Furthermore, even for a given value of $d$, the total
interaction strength $\gamma_{0}+2\gamma_{l}$ (which is proportional
to $c_{0}+2c_{l}$) is not in one to one correspondence with the energy
and therefore cannot be used to characterize the gas. Fig 2(b) shows
that in the regime $d\ll1$, the energy indeed depends only on $\gamma_{eff}$
of (\ref{eq:gam_eff}). The results are consistent with (\ref{eq:c_effg}).

\selectlanguage{american}%
\begin{figure}[H]
\selectlanguage{english}%
(a)\qquad{}\qquad{}\qquad{}\qquad{}\qquad{}\qquad{}\qquad{}(b)

\includegraphics[scale=0.6]{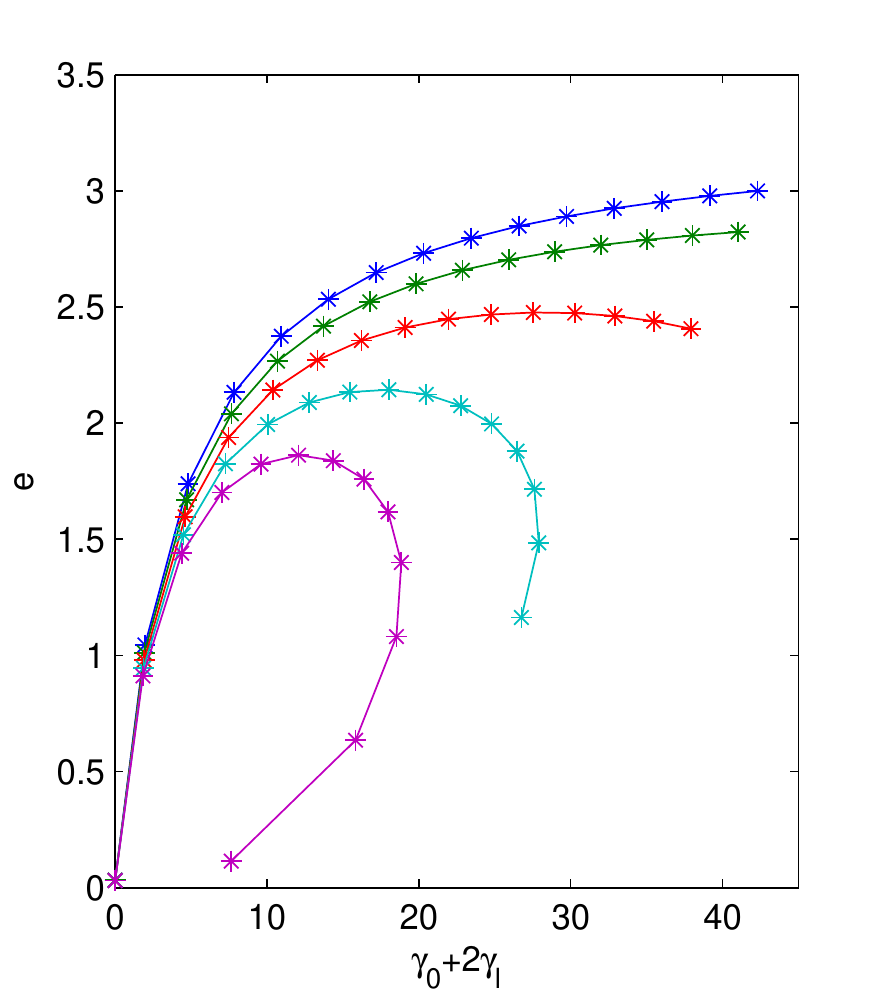}\includegraphics[scale=0.6]{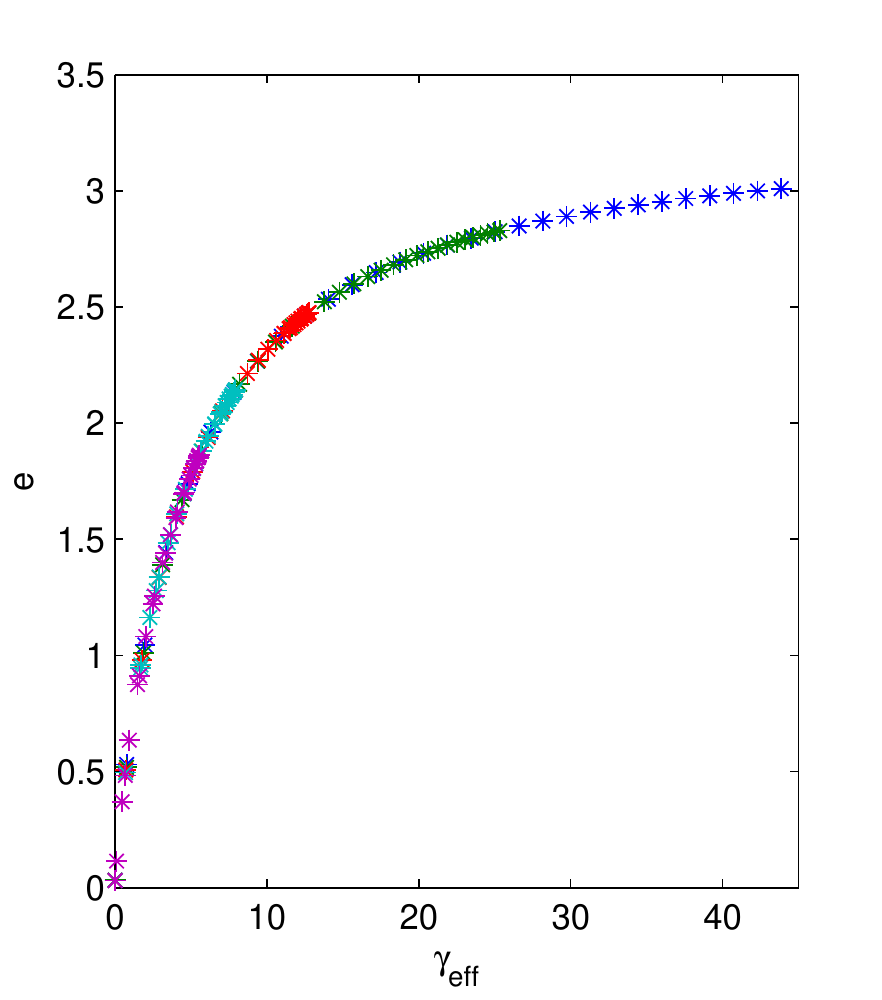}

\selectlanguage{american}%
\caption{\selectlanguage{english}%
(Color online) The dimensionless energy $e$ of (\ref{eq:dle}) as
a function of dimensionless interaction strengths for $c_{l}=-c_{0}/4$
(namely, $r=-0.25$) and $0<\alpha_{0}<30$. (a) $e$ as a function
of $\gamma_{0}+2\gamma_{l}$. Different lines represent different
choices of $d$ of (\ref{eq:dim_less}), from top to bottom: $d=0$
(blue), $d=0.02$ (green), $d=0.04$ (red), $d=0.06$ (turquoise),
$d=0.08$ (purple). Points where the effective interaction is attractive
were excluded from the figure (these were supposed to appear in the
bottom purple curve in the regime $x>x_{0}=2$, see Eq. (\ref{eq:x_0})),
so that the highest value of $x$ which does appear in the figure
is $x=1.94$ and it corresponds to the purple point $\left(7.7,0.114\right)$.
(b) The energy $e$ of (a), plotted as a function of $\gamma_{eff}$
(Eq. (\ref{eq:gam_eff})).\selectlanguage{american}%
}
\end{figure}

\selectlanguage{english}%

\section{Regime of stability and definition of dilute gas}

The Bethe ansatz equations (\ref{eq:betN}) and the effective interaction
(\ref{eq:c_eff}), are valid only where (\ref{eq:hilazon}) is satisfied.
Therefore, it is important to identify the regime where $\left(k_{j}-k_{s}\right)l\ll1$.
In the original LL model, the ground state energy and the values of
$k$'s are maximal for strong interactions, $c\rightarrow\infty$,
where $k_{n}=\frac{2\pi}{L}n$ and $n$'s are integers $n=-\frac{N}{2},\ldots,\frac{N}{2}$.
Then, the maximal absolute value of $k$ is $k_{max}=\frac{\pi N}{L}$
and for all $j,s$, 
\begin{equation}
\left(k_{j}-k_{s}\right)l<\frac{2\pi Nl}{L}.
\end{equation}
For dilute gas, the inter-particle separation $L/N$ is much larger
then the interaction range $l$ and (\ref{eq:hilazon}) is satisfied.

The same argument can be written for the three $\delta$-functions
interaction potential (\ref{eq:3D_shrod}). If $\theta$ of (\ref{eq:theta})
is a monotonic increasing function of $k$, the ground state is given
by $n_{j}=j$, $\left(j=1,\ldots,N\right)$ and $k_{max}=k_{N}$.

Let us analyze the function $\theta\left(k\right)$ and identify the
regime of parameters where it is monotonic. first, note that $\theta\left(k\right)$
is monotonically increasing if and only if $f\left(k\right)\equiv\frac{1}{i}\left[Q_{0}+2Q_{l}\cos\left(kl/2\right)\right]$
is monotonically increasing. For $k\rightarrow0$, $f\left(k\right)=-\frac{m}{\hbar^{2}k}c_{eff}$
and therefore it is monotonically increasing as long as $c_{eff}>0$.
Hence, if $c_{eff}>0$, there exist some $k^{*}$ (which depends on
the parameters $c_{0},c_{l},l$ and does not depend on $L$ and $N$
since $\theta$ is independent of these variables) such that for all
$k<k^{*}$, $\theta\left(k\right)$ is monotonically increasing. For
the ground state, the states with the smallest $\left|k_{j}\right|$
are occupied, namely, $n_{j}=j$ with $j=1,\ldots,N,$ and 
\begin{equation}
G\left(k_{j}\right)<\pi N.
\end{equation}
$\theta$ is an angle variable and therefore it is bounded (actually,
for very small $k$, $\theta=-\pi$). Hence
\begin{equation}
\left|k_{j}\right|<\mathrm{\left(const\right)}\frac{N}{L}
\end{equation}
and can be made arbitrary small. Now, by increasing $L$ (or decreasing
$N$), one may tune the value of $k_{max}$ such that the conditions
\[
k_{max}<k^{*}
\]
and
\[
k_{max}l\ll1
\]
are satisfied simultaneously, the regime (\ref{eq:hilazon}) of dilute
gas is achieved and our solution is correct up to a term of order
$Nl/L$. 

For a dilute gas there is a range of parameters where $c_{eff}>0$
and the solution is stable. There is also a range of parameters where
$c_{eff}<0$ and the system is unstable.

\section{Summary and discussion}

In this paper, we analyzed a one dimensional dilute Bose gas for an
extension of the LL model defined by (\ref{eq:3D_shrod}). By dilute
gas, we mean that $l\ll L/N$, that is, the effective size of a particle
$l$ (for example, the Van-der-Waals radius of an atom) is much smaller
than the inter-particle distance. Using this assumption and the Bethe
ansatz, we derived the approximate equations for the spectrum (\ref{eq:bet12})
,(\ref{eq:bet21}), (\ref{eq:betN}). In principle, these can be solved
numerically. For low energies in this situation $\left|k_{j}l\right|\ll1$
and the model can be approximated by a LL model with one $\delta$-function
of strength $c_{eff}$ given by (\ref{eq:c_eff}) and in dimensionless
units by (\ref{eq:gam_eff}). The error of this approximation is of
order $Nl/L$. This is a good approximation for the dilute gas. The
effective interaction $c_{eff}$ depends on $c_{l}$ and $c_{0}$
but also on the ratios between the characteristic potential energy
scales, $c_{l}/l$ and $c_{0}/l$, and the kinetic energy scale, $\hbar^{2}/ml^{2}$,
of a particle trapped in a well of length $l$.

Naively one would expect that for small $k_{j}$, $c_{eff}\approx c_{0}+2c_{l}$.
It turns out to be correct for relatively weak interaction energy.
For stronger interactions, $c_{eff}$ becomes very small and even
changes its sign (see Fig. 1). Note that this result holds also for
the two particle case where it is exact. It is a surprising result,
verified numerically in Fig. 2 and its experimental verification should
be considered a challenge. The knowledge of $c_{eff}$ enables to
calculate the ground state and the low excited states if the conditions
for stability are satisfied. In section IV, the ground state is calculated
in the thermodynamic limit for a dilute gas. In particular, it is
demonstrated to depend on all parameters via $c_{eff}$. We have shown
that for a dilute gas there is a regime of parameters where $c_{eff}>0$
and therefore the system is stable. For other parameters, $c_{eff}<0$
and the dilute gas is unstable. In this regime, the results of \cite{beta_prl_corr,beta_stat_mech}
regarding dynamics of attractive gas might be realized. If the gas
is not dilute, we cannot determine the stability of the system. This
theoretical model enables to predict qualitative features of interacting
bosons for realistic systems.

The potential (\ref{eq:3D_shrod}) can be realized, for example, in
optical lattices \cite{Bloch_tonks} with tight harmonic trapping
along two perpendicular directions ($E\ll\hbar\omega_{\perp}$) and
almost flat potential along the third direction. The inter-particle
interactions are in three dimensions and can be modeled by a ``delta
shell'' potential\foreignlanguage{american}{
\begin{equation}
V\left(r\right)=\left\{ \begin{array}{cll}
\frac{3c_{0}\hbar}{4r_{in}^{3}m\omega_{\perp}}\qquad & for\qquad & r<r_{in}\\
\frac{c_{l}\hbar}{2r_{out}^{2}\varepsilon_{out}m\omega_{\perp}}\qquad & for\qquad & r_{out}<r<r_{out}+\varepsilon_{out}\\
0 &  & \mathrm{otherwise}
\end{array}\right..\label{eq:delta shell}
\end{equation}
Where $r_{in},\varepsilon_{out}\rightarrow0$. In a previous work
\cite{Ketterle&us}, we calculated the scattering length $a$ and
the effective range $r_{e}$ of the potential (\ref{eq:delta shell})
(see App.A of \cite{Ketterle&us}), wrote a three dimensional Schr}$\ddot{\text{o}}$\foreignlanguage{american}{dinger
equation and integrated it over two axes to obtain a one dimensional
equation of the form }(\ref{eq:3D_shrod}). This leads to the relations
\begin{equation}
r_{out}=\sqrt{3}l/2,
\end{equation}
\begin{equation}
a=\frac{1}{4\hbar\omega_{\perp}}\left(c_{0}+2c_{l}\right)\label{eq:a}
\end{equation}
and
\begin{equation}
r_{e}=\frac{2c_{l}l^{2}}{a\left(c_{0}+2c_{l}\right)}+\frac{2a}{3}.
\end{equation}
As seen from (\ref{eq:c_eff}), for $l=0$, $c_{eff}$ is proportional
to the scattering length. However, for $l\neq0$, $c_{eff}$ cannot
be expressed in terms of $a$ and $r_{e}$. Therefore, it motivates
introducing an effective scattering length that dominates the spectrum.

From an experimental point of view, it looks that $c_{eff}$ is the
only quantity that one can measure in order to characterize the inter-particle
interactions (because it determines the spectrum). Hence, it makes
sense to define an effective scattering length
\begin{equation}
a_{eff}=\frac{c_{eff}}{4\hbar\omega_{\perp}}.\label{eq:a_eff}
\end{equation}
This scattering length, which includes corrections originating in
the non-vanishing interaction range, is unique for one dimensional
bosonic systems. 
\begin{acknowledgments}
We thank Eliot Lieb and Avy Soffer for suspecting an error in the
original version of the work and Daniel Podolsky, Yoav Sagi and Efrat
Shimshoni for illuminating and informative discussions. The work was
supported in part \foreignlanguage{american}{by the Israel Science
Foundation (ISF) grant number 1028/12, by the US-Israel Binational
Science Foundation (BSF) grant number 2010132 and by the Shlomo Kaplansky
academic chair.}
\end{acknowledgments}
\selectlanguage{american}%
\bibliographystyle{h-physrev}
\addcontentsline{toc}{section}{\refname}\bibliography{C:/Users/Hagar/Dropbox/harmonic_paper/referances}

\begin{thebibliography}{10}

\bibitem{Pethick@Smith}
C.~Pethick and H.~Smith,
\newblock {\em Bose-Einstein Condensations in Dilute Gases} (Cambridge
  University Press, 2002).

\bibitem{Pita_book}
L.~Pitaevskii and S.~Stringari,
\newblock {\em Bose-Einstein Condensation} (Oxford science publications, 2003).

\bibitem{PS_review}
F.~Dalfovo, S.~Giorgini, P.~Pitaevskii, Lev, and S.~Stringari,
\newblock Rev.Mod.Phys {\bf 71}, 463 (1999).

\bibitem{Girardeau}
M.~Girardeau,
\newblock J. Math. Phys. {\bf 1}, 516 (1960).

\bibitem{Olshanii}
M.~Olshanii,
\newblock Phys. Rev. Lett. {\bf 81}, 938 (1998).

\bibitem{Bloch_tonks}
B.~Paredes {\em et~al.},
\newblock Nature {\bf 429}, 277 (2004).

\bibitem{beta_science}
T.~Kinoshita, T.~Wenger, and D.~S. Weiss,
\newblock Science {\bf 305}, 1125 (2004).

\bibitem{beta_prl_exp}
F.~Meinert {\em et~al.},
\newblock Phys. Rev. Lett. {\bf 115}, 085301 (2015).

\bibitem{Lieb}
E.~H. Lieb and W.~Liniger,
\newblock Phys. Rev. {\bf 130}, 1605 (1963).

\bibitem{bog_book}
V.~E. Korepin, N.~M. Bogoliubov, and A.~G. Izergin,
\newblock {\em Quantum Inverse Scattering Method} (Cambridge University Press,
  1997).

\bibitem{Tka_book}
M.~Takahashi,
\newblock {\em Thermodynamics of one-dimensional solvable model} (Cambridge
  University Press, 2005).

\bibitem{Liev_Spect}
E.~H. Lieb,
\newblock Phys. Rev. {\bf 130}, 1616 (1963).

\bibitem{zoran}
Z.~Ristivojevic,
\newblock Phys. Rev. Lett. {\bf 113}, 015301 (2014).

\bibitem{3_Del_long}
T.~Cheon and T.~Shigehara,
\newblock Phys. Lett. A. {\bf 243}, 111 (1998).

\bibitem{3_del_PRL}
T.~Cheon and T.~Shigehara,
\newblock Phys. Rev. Lett. {\bf 82}, 2536 (1999).

\bibitem{Ketterle&us}
H.~Veksler, S.~Fishman, and W.~Ketterle,
\newblock Phys. Rev. A. {\bf 90}, 023620 (2014).

\bibitem{beta_prl_corr}
P.~Calabrese and J.-S. Caux,
\newblock Phys. Rev. Lett. {\bf 98}, 150403 (2007).

\bibitem{beta_stat_mech}
P.~Calabrese and J.-S. Caux,
\newblock J. Stat. Mech. , P08032 (2007).

\end{thebibliography}
\selectlanguage{english}%

\end{document}